\documentstyle[aaspp,11pt]{article}
\newcommand{\be} {\begin{equation}}

\newcommand{\ee} {\end{equation}}
\newcommand{\co}{\rm}

\newcommand{\bc}{\begin{center}}
\newcommand{\ec}{\end{center}}

\begin{document}

\title{\large{\bf A NEW TECHNIQUE FOR THE DETECTION OF PERIODIC SIGNALS
 IN ``COLOURED" POWER SPECTRA }}

\author{\bf G. L. Israel\altaffilmark{1}}
\affil{International School for Advanced Studies (SISSA--ISAS), V. Beirut
2--4, I--34014 Trieste, Italy} 
\affil{Dipartimento di Fisica, Universit\`a di Roma ``La Sapienza'', P.le A. 
Moro 2, I--00185, Roma, Italy}
\affil{e--mail: gianluca@{\co vega.sissa.it}}

\and

\author{\bf L. Stella\altaffilmark{1, 2} }
\affil{Osservatorio Astronomico di Brera, Via E. Bianchi 46,
I--22055 Merate (Lc), Italy}
\affil{e--mail: stella@{\co antares.merate}.mi.astro.it}

\altaffiltext{1}{Affiliated to the International Center for Relativistic 
Astrophysics}
\altaffiltext{2}{{\co Now at the Astronomical Observatory of Rome, Via 
dell'Osservatorio 2, I--00040, Monteporzio Catone (Roma), Italy}}

\begin{center}
Accepted for publication in {\it The Astrophysical Journal}
\end{center}

\begin{abstract}

The light curves from a variety of celestial objects display aperiodic
variations, often giving rise to red--noise components in their power spectra. 
Searching for a narrow power spectrum peak resulting from 
a periodic modulation over the frequency range in which these
``coloured'' noise components are dominant has proven a very complex task. 
Commonly used methods rely upon spectral smoothing or incoherent 
summation of sample spectra in order to decrease the variance of the 
power estimates. The consequent reduction in frequency resolution 
causes also a reduction of sensitivity to periodic signals.

We develop here a technique aimed at detecting periodicities in the
presence of ``coloured" power spectrum components, 
while mantaining the highest Fourier frequency resolution.
First we {\co introduce a simple approximation to the} statistical properties 
of the ``coloured'' power spectra from celestial objects, based on a 
few examples and the theory of linear processes. 
We then estimate the continuum components in the power spectrum  
through an {\it ad hoc} smoothing technique. This involves
averaging the spectral estimates adjacent to each frequency over a 
{\co suitably chosen}  
interval, in order to follow steep red--noise features and 
produce estimates that are locally unaffected by the possible presence of a 
sharp peak. By dividing the sample spectrum by the smoothed one, a white--noise 
like spectrum is obtained, the approximate probability distribution  
of which is derived.
A search for coherent pulsations is then carried out by looking for peaks
in the divided spectrum,  
the chance probability of which is below a given detection threshold. If
no significant peaks are found, an upper limit to the amplitude of a
sinusoidal modulation is worked out for each searched frequency.

The technique is tested and its range of applicability determined through
extensive numerical simulations. We present also an application to the 
X--ray light curves of V$0332$$+$$53$, a highly variable accreting X--ray 
pulsar, and GX$13$$+$$1$, a bright and variable accreting source in the 
galactic bulge. 

\end{abstract}

\keywords{ Numerical methods --- Pulsation --- Stars: Neutron --- Pulsars 
 --- Oscillations --- X--ray: Binaries} 
\newpage

\section{INTRODUCTION}
Since the prehistorical efforts aimed at developing the calendar,
the detection and investigation of periodic phenomena
has played a major role in astronomy.
Crucial information is obtained through the observation and measurement
of periodicities in many classes of celestial bodies encompassing all scales
from comets and asteroids to the largest structures currently known at
cosmological distances. In some cases periodic signals from the cosmos
can be measured to an exceptionally high accuracy, that rivals that of
atomic clocks (Kaspi, Taylor \& Ryba 1995).
Astronomical time series of increased statistical quality, time resolution
and duration have become available over the last two decades for a variety
of objects and over different bands of the electromagnetic spectrum.
Power spectrum analysis is probably the single most important technique that
is applied to these series in order to: (a) detect periodicities (or
quasi--periodicities) by the presence of significant power spetrum peaks; 
(b) characterise the noise variability through the study of
continuum power spectrum components. In particular, recent applications
to high energy astronomical time series have been especially
numerous and successfull, as a consequence of the
pronounced  variability (often both periodic and aperiodic in nature)
detected in many sources and the availability of long uninterrupted
observations  (up to several days) of high signal to noise ratio.
The continuum  power spectral components arising from noise variability
usually increase towards  lower frequencies ({\it red noise}), often  in a
power law--like fashion.  Their study has proven to be a useful tool for
morphological classifications and, sometimes, has provided  constraints on
physical models (e.g. Stella 1988;
Hasinger \& van der Klis 1989; van der Klis 1995).
The periodic modulations revealed in a number of high energy
sources often arise from the rotation of compact magnetic
stars, or the orbital motion of a binary system. The detection and
accurate  measurement of these periods provides a tool of paramount
importance.
A variety of other periodic or quasi--periodic phenomena in X--ray
sources have been discovered  over a range of timescales (from tens of
milliseconds to years).

Astronomical observations rely more and more upon
photon counting instruments; therefore, measurement errors are often
dominated by the statistical uncertainties originating from the Poisson
distribution of the counts. This translates into a
white noise power spectrum component of known amplitude, which, after
normalisation, follows a $\chi^2$ distribution with 2 degrees of freedom
($\chi^2_2$).
Any intrinsic variability of the source, either resulting
from periodic signal(s) or from noise(s), must possess significant power
above the counting statistics white noise component in order to be detected.

Traditionally, the detection of periodic signals through peaks
in the sample spectrum has been carried out either (i) by eye,
in all those  cases in which the peak amplitude is so large that it is
self--evident or (ii) by ruling out (at a given  confidence level) that a
peak originates from an underlying white noise.
The latter technique implicitely assumes that the power spectra do not possess
any conspicuous ``coloured" component above the
white noise. As mentioned above, this hypothesis
is not verified at least over a range of frequencies
in many instances. Indeed the very
presence of ``coloured" continuum power spectrum components resulting from 
source variability noise makes the detection of significant power
spectrum peaks a difficult statistical problem.
In general, establishing whether or not a sample spectrum peak
originates from a periodic modulation requires 
an evaluation of the peak significance with respect to the
{\it local} continuum of the sample spectrum which, in turn, can be dominated
by the   aperiodic variability of the source.
Techniques along these lines have been developed, which often  
rely upon smoothing or incoherent summing of sample spectra
to decrease the variance of the power estimates and/or allow the use of
relatively simple statistics. In this way, however,
the frequency resolution and, correspondingly, the
sensitivity of the searches is reduced
(e.g. Jenkins \& Watts 1968; van der Klis {\co 1988}).
Moreover, standard spectral smoothing does not allow to reproduce power
law--like spectral shapes with acceptable accuracy.

This paper describes a new technique for detecting  power spectrum peaks
arising from  a periodic modulation, in the presence of ``coloured" power
spectrum components, while preserving the Fourier frequency resolution.
In this technique, the continuum components of the 
spectrum at the $j$--th frequency are 
evaluated based on an {\it ad hoc} smoothing technique which involves
averaging the spectral estimates adjacent to $j$--th frequency over a given
logarithmic interval excluding the $j$--th frequency itself. 
The advantage of this type of
smoothing is that, while it allows the continuum features of the power
spectrum to be followed, it is locally unaffected (i.e.  for the same
Fourier frequency) by the presence of sharp power spectrum peaks. By
dividing the sample spectrum by the smoothed one a white--noise like
spectrum is obtained, the approximate probability distribution function 
($pdf$) of which is derived
based on the characteristics of the sample spectrum.
A search for coherent pulsations is then carried out by looking for peaks
in the divided spectrum, for which
the probability of chance occurrence is below a given detection level. If
no significant peaks are found, an upper limit to the amplitude of a
sinusoidal modulation is worked out for each searched frequency.

Our treatment assumes that the instrumental noise is due to Poisson 
statistics; as such it can be readily applied to observations with photon
counting detectors in any band of the electromagnetic spectrum. The
generalisation to the case of a Gaussian instrumental noise
is straightforward.
The time series are supposed to  be equispaced and continuous.

The paper is structured as follows: in section 2
we {\co introduce a simple approximation to} the $pdf$ of the power 
estimates in the sample spectrum of cosmic 
sources characterised by ``coloured" noise.
Section 3 describes the smoothing algorithm that we devised in order to 
estimate the corresponding continuum power 
spectrum components, even in the presence of quite steep red--noises. 
In section 4 we derive the $pdf$ of the white--noise like spectrum that 
is obtained by dividing the sample spectrum by the smoothed one. 
The prescription for detecting significant periodic signals and deriving their 
amplitude, is given in section 5.
This includes also how to obtain upper limits in the case in which 
no significant peak is found. Section 3--5 summarise also the results from 
the extensive numerical simulations that were carried out in order to 
assess the reliability of the technique. 
In section 6 an application to the ``coloured" power spectra from the 
X--ray light curves of an accreting X--ray pulsar (V0332+53) and a 
bright galactic bulge source (GX13+1) is presented. Our conclusions are in 
section 7.

\section{THE DISTRIBUTION OF THE POWER ESTIMATES IN THE SAMPLE SPECTRUM}

When ``coloured" continuum power spectrum components resulting from
source  variability are present, the statistical distribution of the
corresponding power estimates cannot be derived in general from first
principles. In the presence of extensive and repeated observations, the
statistical  properties of these components could be
obtained directly from the data. In practice this is difficult to do,
because of the limited duration of the observation and the
characteristic red--noise spectra that are commonly found.
{\co An additional limitation derives from the fact that many
cosmic sources display different activity states, often characterised by 
different luminosity and/or energy spectrum properties. 
A given activity state can last for 
time intervals as short as minutes; in same cases this imposes the 
tightest constraint on the lowest frequencies that can be studied in the 
sample spectrum, without violating the hypothesis of stationarity.}

A single sample spectrum is often calculated over the
entire observation duration, $T$, in order to explore the  lowest possible
frequencies, while maintaining the highest Fourier resolution
($\Delta\nu_F = 1/T$). In this
case only one power estimate is obtained for each Fourier frequency and the
statistical distribution of the noise component(s) from the source remains
unexplored. 

Alternatively  the observation can be divided in a series of $M$ consecutive
intervals and  the distribution of the power estimates investigated over the
ensemble of the sample spectra from individual intervals of duration $T/M$.
To illustrate this, we analysed a $\sim$5.5~hr long observation 
of the accreting black hole candidate Cyg~X--1 {\co in its so--called ``low 
state''}, one of the most 
variable X--ray binary sources in the sky.  
The 7.8~ms resolved X--ray light curve (1--20 keV energy range) was divided 
into $M=1244$ intervals of 16~s and a sample spectrum calculated for each 
interval. The sample spectrum obtained by averaging these $M$ spectra
is given in fig.~1a.  
Fig.~1b shows the distribution of the power estimates for
selected frequencies over the $M$ sample spectra. Each distribution is
normalised by $2/<P_j>$,  where $<P_j>$ is the estimate of the average power at
the $j$--th Fourier  frequency $\nu_j$. It is apparent that in all cases the
distribution is close to a $\chi^2_2$  $pdf$ (also plotted for comparison). 
A Kolmogorov--Smirnov test
gives a probability of $\sim$$20$--$60$\% that the observed distributions
result from a $\chi^2$ $pdf$. {\co Similar results were obtained for 
the sample spectra from the light curves of a few other accreting 
compact stars in X-ray binaries}. 
By extrapolating these results, we assume that, {\co (for a given 
activity state)}, the ``coloured"
noise components in the sample spectra of cosmic sources also follow a rescaled
$\chi^2_2$--distribution\footnote{\co Some caution is necessary for red noise 
spectra with a power law slope steeper than -2. In these cases the source 
variability on timescales comparable to those over which the sample 
spectrum is calculated, can cause a substantial low-frequency leakage, which in 
turn might alter the distribution of the power estimates.  
To limit the effects of this leakage our technique includes the possibility 
of subtracting polynomial trends from the light curves (see Deeter 1984).} 
{\co (see also van der Klis 1988)}. 

There is at least a very important class of random processes for which the
power  spectral estimates possess properties compatible with those discussed
above. These are linear processes, $y(t)$,  in which a white noise, $z(t)$, is
passed through a linear filter $h(t)$, i.e.
\be
y(t)-\mu=\int_{0}^{\infty}h(\tau)z(t-\tau)~d\tau~,
\label{eq:lp1}
\ee
\noindent
where $\mu=E[y(t)]$ is the mean of $y(t)$ and $E[z(t)]=0$.
The power spectrum, $\Gamma_y$ of a linear process is given by:
\be
\Gamma_y(\nu)=|H(\nu)|^2 \Gamma_z(\nu)~,
\label{eq:py}
\ee
where $\Gamma_z$  is the power spectrum of $z(t)$ and
$H(\nu)$ is the frequency response function of the linear filter $h(t)$.
The power spectrum is the average over the realisations of the
sample spectrum, i.e. $\Gamma_y(\nu)= E[Y(\nu)]$ and 
$\Gamma_z(\nu)= E[Z(\nu)]=2$. The latest equality implies that the sample 
spectrum of the input white noise is normalised such as to follow a 
$\chi^2_2$ {\it pdf}
(e.g. Jenkins \& Watts 1968).
Given a white noise source and a suitable linear filter it is then possible to
generate a random process with arbitrary spectrum.
In particular, it follows from eq.~2 that for a given frequency $\nu$,
the sample spectrum, $Y(\nu)$, of the linear
process follows the same $\chi^2_2$ distribution of 
the sample spectrum of the input white noise $Z(\nu)$, except for a rescaling  
factor of $|H(\nu)|^2$.
Therefore, the $pdf$ of $Y(\nu)$ is
\begin{equation}f_{\nu}(y)
=\frac{e^{-y/2|H(\nu)|^{2}}}{2|H(\nu)|^{2}}~(y) ~.
\label{eq:pdfeq}
\end{equation}
{\co Based on the discussion above we adopt linear processes to model the 
sample power spectra (and their $pdf$) resulting from ``coloured" noise 
variability of cosmic sources.} 

In practical applications the sample spectrum of astronomical time series
$P_j$ comprises a white noise component resulting from measurement
uncertainties (Poisson noise in the case of photon counting detectors). The
power estimates of the white noise resulting from Poisson statistics are
distributed according  to a $\chi^2_2$ $pdf$, if the normalisation  
\be
P_{j}=\frac{2}{N_{\gamma}}|a_{j}|^{2}
\ee
is adopted, where $N_{\gamma}$ is the total number of photons in the light
curve and $a_{j}$ the complex Fourier amplitudes (see e.g. Leahy et al. 1983). 
In the case of a Gaussian instrumental noise with mean zero and
variance  $\sigma^2$, $N_{\gamma}$ {\co is to be} replaced by $N\sigma^2$, where
$N$ is the number of points in the light curve. Therefore in the regions of the
sample spectrum which are dominated by instrumental (white) noise, 
the power estimates will follow a $\chi^2_2$ $pdf$. 
We assume that this instrumental white noise component
can be interpreted as the input process $z(t)$, such that
eq.~(\ref{eq:lp1})
still holds. While clearly  non--physical, this assumption involves no
(statistical) approximation and allows to considerably simplify
our treatment.
In particular 
it follows that if the square modulus of the frequency response function 
$|H(\nu)|^2$ were known, then multiplying
eq.~(\ref{eq:py}) by  $|H(\nu)|^{-2}$ the spectrum of the (instrumental) white
noise would be recovered. In this case 
the search for significant power spectrum peaks
arising from a periodic signal could be carried out by using standard
techniques.
In practice $|H(\nu)|^2$ must be estimated through the sample spectrum.
One possibility  would be to model the power spectrum continuum components
by adopting an appropriate maximum likelihood technique ({\co 
Anderson, Duvall \& Jeffries 1990; Stella et al. 1996;} Arlandi et al.
1996) and use the best fit function to estimate $|H(\nu)|^2$. This approach,
however, faces difficulties with the subjective choice of the model function
and,  more crucially, the estimate of the statistical uncertainties of the best
fit at any given frequency.
Therefore we prefer to evaluate $|H(\nu)|^2$ through a suitable smoothing
algorithm as described in sect. 3. 

The discussion above concentrates on individual sample spectra. It is not
uncommon, however, that the sample spectrum is obtained from 
the sum of the sample spectra of $M$ different intervals. In
general, therefore, each spectral estimate can be the sum of 
$M$ individual estimates and its statistical distribution is, therefore,
related to a $\chi^2$--distribution with $2M$ degrees of freedom ($dof$). This
case is described in detail in the Appendix. 

\section{{\co THE} SMOOTHING ALGORITHM}

As the goal of any periodicity search
is to detect a sharp peak over the underlying sample spectrum continuum,
the power in a (possible) 
peak should not affect the estimate of the continuum
(otherwise the sensitivity of the search would be reduced). 
This implies that for each frequency $\nu_j$ the continuum should be 
estimated through an interpolation of the
sample spectrum which excludes $P_j$ itself and uses  the power
estimates over a range of nearby frequencies at the left and right of
$\nu_j$. In the language of the smoothing functions, this corresponds to
a well--known class of spectral windows which are zero--valued at the central
frequency. 
We adopt for semplicity a rectangular window (with a central gap)
that extends over a total of $I$ Fourier frequencies, giving a
width of $\Delta\nu_{tot} = I \Delta\nu_{F}$.

In conventional  smoothing
$(I-1)/2$  Fourier frequencies are used shortwards and longwards
of the central frequency $\nu_j$, such that the same smoothing width
$\Delta\nu_{left}=\Delta\nu_{right}= (I-1) \Delta\nu_{F}/2$ is obtained on both
sides of the central frequency.  The problem with this kind of smoothing
is that it does not approximate with acceptable accuracy the steep power--law
like red--noise components that are often found in the sample spectra 
of cosmic sources. 
Fig.~2 shows the results of 100 simulations of {\co three} different types of
red power spectra consisting of: a Lorentzian centered at zero
frequency  (spectrum A), a power--law  with a slope of  $-1.5$
(spectrum B) {\co and a power law with a slope of $-2$ (spectrum C)}. 
In {\co all} cases a  quasi--periodic oscillation broad peak centered
around 100 Hz was included, together with a counting statistics white
noise component. A smoothing width of $I=30$ Fourier frequencies was used.
It is apparent that while conventional smoothing
(dotted lines in fig.~2) reproduces fairly accurately the
characteristics of spectrum A,  it fails to reproduce the steep
decay from the lowest frequencies of spectrum B {\co and C}.  Moreover edge 
effects dominate the estimate of the smoothed spectrum for the first 
$(I-1)/2$ frequencies.

A far better result is obtained if the smoothing over $I$ Fourier frequencies
is distributed such that its logarithmic frequency width is (approximately) the
same on both sides of $\nu_j$, i.e.
$\log(\Delta\nu_{j,left})=\log(\Delta\nu_{j,right})$.
This approach builds on the obvious fact that a power law 
spectrum is a straight line in a log--log representation.
Considering that $\Delta\nu_{tot}$=$\Delta\nu_{j,left}+\Delta \nu_{j,right}$ 
it follows that
\be
\left\{ \begin{array}{ll}
          \Delta \nu_{j,left} = \Delta \nu_{tot} - \Delta \nu_{j,right} \\
          \Delta\nu_{j,right} = \displaystyle{\frac {\left[ -(2 \nu_j -
          \Delta \nu_{tot}) + (4\nu_j^{2} + \Delta \nu^{2}_{tot})^{1/2} 
          \right]} {2}}~~~~~.
        \end{array}
\right.
\ee
In this scheme the smoothed spectrum $S_j(I)$, 
that we adopt as the estimator of 2$|H(\nu_j)|^2$, is calculated as follows
\be
 S_j(I) = \frac{1}{2I_{j,left}}
  \sum_{i=j-I_{j,left}}^{j-1}
 P_i+{\frac{1}{2I_{j,right}}} \sum_{i=j+1}^{j+I_{j,right}} P_i ~~~~,
\label{eq:s1}
\ee
where $I_{j,left}$ and $I_{j,right}$ (rounded to the nearest integers)
are the number of Fourier frequencies 
in $\Delta\nu_{j,left}$ and $\Delta\nu_{j,right}$, respectively.
\be
\sigma^2_{S_j(I)} = {\frac{1}{2}}\sqrt{\left[{\frac{1}{I_{j,left}}}
  \sum_{i=j-I_{j,left}}^{j-1} P_i\right]^{2}
  +\left[{\frac{1}{I_{j,right}}} \sum_{i=j+1}^{j+I_{j,right}} P_i\right]^{2}} 
  ~~~~~.
\label{eq:s2}
\ee
By propagating eq.~\ref{eq:s2} we obtain the variance $2P_i$ of the  $P_i$ 
variables over the smoothing  formula (cf. eq.~\ref{eq:s1}) 
The solid lines in fig.~2 show the estimate of the continuum power
spectrum components (and therefore of 2$|H(\nu_j)|^2$) obtained by using the
above technique; it is apparent that also the low-frequency end of
spectra B {\co and C} is reproduced quite well, and that edge effects are 
nearly absent.

In general $I$, the number of Fourier frequencies defining the smoothing width  is
to be adjusted so as to closely follow the sharpest continuous features of the
sample spectrum (samething that favors low values of $I$),  while mantaining the
noise of the smoothed spectrum as low as  possible (something that favors high
values of $I$). To this aim we consider  the divided sample spectrum $R_j(I)=2
P_j/S_j(I)$, i.e. the ratio of the sample spectrum and the smoothed spectrum for a
range of different values of $I$.  If $S_j(I)/2$  provides a reasonably good
estimate of $|H(\nu_j)|^2$, then $R_j(I)$ will approximately follow the
$\chi^2_2$--distribution of the input white noise, at least for relatively small
values of $R_j(I)$  ($\leq 15-20$, see sect.~4). {\co A Kolmogorov--Smirnov (KS)
test  can be used in order to derive out of different trial values of the width
$I$, the one that makes the distribution of  $R_j(I)$ closest to a $\chi^2_2$
$pdf$.} The KS test is especially sensitive to differences away from the
tails of the distributions (see e.g. Press et al. 1992). 

Fig.~3 shows the results from simulations in which the KS probability
is calculated as a function of $I$ for {\co four} different types of spectra
each with 5000 Fourier frequencies. Each point in fig.~3 represents the
average over 100 simulations. The second, third {\co and fourth} 
panels refer to spectra {\co A, B and C} of fig.~2, respectively. In 
{\co all} three 
cases the KS probability shows a broad maximum around values of  $I\sim 100$.
For higher values of $I$ the smoothed spectrum becomes gradually less
accurate in reproducing the shape of the sample spectrum, whereas for lower
values of $I$ the scatter in the estimates of $S_j(I)$  plays an increasingly
important role in distorting the $pdf$ of $R_j(I)$ away from a
$\chi^2_2$--distribution. Note that, as expected, the KS probability
monotonically increases with $I$ in the case of a white noise sample spectrum
(see upper panel of fig.~3).

{\co In the following} we adopt $S_j(I_o)/2$, the smoothed sample 
spectrum with a width 
$I_o$ that maximises the probability of the KS test described above.
In practice values of $I_o$ between 30--40 and the number of Fourier
frequencies in the sample spectrum are to be used (see sect.~4).

\section{THE DISTRIBUTION OF THE DIVIDED SAMPLE SPECTRUM $R_j(I)$}

The smoothed sample spectrum $S_j(I_o)/2$ provides our estimate of
$|H(\nu_j)|^2$, in the sense discussed in the previous section. Therefore 
we adopt the divided spectrum $R_j(I_o) = 2 P_j/S_j(I_o)$ as the estimator of the
white noise spectrum of the input linear process. The search for
coherent periodicities in the data thus translates into the problem of
detecting significant peaks in $R_j(I_o)$. This, in turn, requires a detailed
knowledge of the expected $pdf$ of $R_j(I_o)$, especially for high values.

For each Fourier frequency $\nu_j$, $R_j(I_o)$ is to be regarded as the
ratio of the random variables $P_j$ and $S_j(I_o)$.
$P_j$ is distributed like a $\chi^2_2$ $pdf$ rescaled to an expectation value of  
$|H(\nu_j)|^2$. 
By approximating $|H(\nu_j)|^2$ with $S_j(I_o)/2$ we have:
\be
   f_{P_{j}}(p)= \displaystyle{\frac{e^{-p/S_j(I_o)}}{S_j(I_o)}~(p)} ~~~~~.
\label{eq:chi}
\ee
The distribution of $S_j(I_o)$ is in general a suitable linear combination
of the $I_o-1$ random variables $P_j$ used in the smoothing. These, in turn,
are  distributed like a rescaled $\chi^2_2$ (cf. eq.~8). 
For sufficiently high values of $I_o$, one can appeal to the
central limit theorem and approximate the  distribution of $S_j(I_o)$ with a
Gaussian distribution of mean $S_j(I_o)$  and variance 
$\sigma^2_{S_j(I_o)}$ (cf. eq.~\ref{eq:s1} and \ref{eq:s2}), i.e. 
\be
     f_{S_j(I_o)}(s) = \displaystyle{\frac{1}{\sigma^2_{S_j(I_o)}\sqrt{2 \pi}}
             e^{~\left[ - (s-S_j(I_o))^{2} /2 \sigma^2_{S_j(I_o)}\right]}} 
             ~~~~~.
\label{eq:gau}
\ee
Note that $P_j$ and $S_j(I_o)$ can be
regarded, for any given $j$, as statistically independent variables
(indeed $P_j$ is not used in the computation of $S_j(I_o)$).
In this case the $pdf$ of $R_j(I_o)$ can be written as
(e.g. Mood, Graybill \& Boes 1974)
\begin{eqnarray}
   f_{R_j(I_o)}(r) &=& \int_{-\infty}^{+\infty} |s| f_{P_j,S_j(I_o)}
   (rs,s) ds  \nonumber \\
   &=& \frac{1}{2 S_j(I_o)\sigma_{S_j(I_o)} \sqrt{2\pi}}~\cdot 
   \int_{0}^{+\infty} s~
   exp \left[- \frac{(s-\mu)^2}{2\sigma^2_{S_j(I_o)}} -
   \frac{rs}{S_j(I_o)} \right] ds ~~~~~, 
\label{eq:fu}
\end{eqnarray}
where we have used the fact that the joint $pdf$ $f_{P_j,S_j(I_o))}(p,s)$ 
is given by the product of $f_{P_j}(p)$ and $f_{S_j(I_o)}(s)$. 

To check the accuracy and range of applicability of the $pdf$ in eq.~
\ref{eq:fu}, we carried out extensive numerical simulations.
Figure~4 shows the results from $2 \times 10^4$ simulations of white noise
sample spectra each cointaing 5000 Fourier frequencies 
({\it i.e.} a statistics of 10$^8$ points). The observed 
distribution of the $R_j(I_o)$ is shown together with the expected $pdf$
derived above. (The $pdf$ in eq.~\ref{eq:fu} and its cumulative
distribution were evaluated numerically through Gaussian integration
routines). In order to the avoid large values of $\sigma_{S_j(I_o)}$ 
arising from small values of $I_{left}$
or $I_{right}$, respectively close to the low-frequency and the 
high-frequency end of the sample spectrum (see eq.~\ref{eq:s2}),
only the powers corresponding to the Fourier frequencies
from j=6 to j=4995 were considered. The
simulations were repeated for different choices of the smoothing width
$I$. It is apparent that while in the cases $I=50$ and $40$ the
$pdf$ in eq.~\ref{eq:fu} provides a very good approximation of the observed
distribution, for $I=30$ and, especially, $I=20$ the expected $pdf$
shows a significant excess for values of $R_j(I)$ larger than $20-30$. This 
effect is due
to the fact that the low--value end of the Gaussian approximation for the $pdf$
of $S_j(I)$ becomes increasingly inaccurate as $I$ decreases.
Therefore in most practical application it is best to use {\co $I \geq 40$}.
On the other hand, being in excess of the observed distribution, the
$pdf$ in  eq.~\ref{eq:fu} would artificially decrease the
sensitivity of searches for significant power spectrum peaks
when used with $I \simeq 20-30 $, but would not favor
the  detection of spurious peaks.

We also tested the reliability of our approximations for the first few
Fourier frequencies of the power spectra (where  $I_{left}\ll I_{right}$)
and the frequencies close to the
Nyquist frequency (where  $I_{left} \gg I_{right}$).
To this aim we carried out $10^6$ simulations of 1000 Fourier frequencies
power spectra from a white noise process, and concentrated on the distribution
of $R_j(I_o)$ for $I_o=100$ and selected values of $j$. Fig.~5 and 6 shows a
comparison of the sample and expected distributions for $j=5,6,7$ and $10$ and
for $j=990, 993, 994$ and $995$. The results clearly show that, for the $pdf$ 
in eq.~10 to provide a good approximation to the simulated distributions
it is necessary to exclude the first and the last $5-6$ frequencies of the
power spectra.
The results above were also determined to be insensitive to the value of $I_o$,
as long as $I_o \geq 40$.

Finally $2 \times ~10^4$ simulations of 5000 Fourier frequencies
sample spectra were carried out for the red
noise spectrum B of sect.~3. The simulated and  predicted $pdf$ of
$R_j(I)$ are shown in fig.~7 for $I = 30, 100, 200$ and $500$. The
values of $R_j(I)$ corresponding to the first and last five Fourier frequencies
were excluded from the distributions. Unlike the white
noise simulations, the width $I$ is here crucial in determining whether
or not the smoothed spectrum closely follows the continuum features of the
sample spectrum. It is seen that the expected $pdf$ closely
follows the simulated distribution for $I = 30$ and {\it I} = 100, whereas
for  $I = 200$ and, especially, $I=500$ the
occurence of high values of $R_j(I)$ {\co is systematically in excess of the 
expected $pdf$}. The latter effect is
clearly due to the fact that, for high values of $I$,
$S_j(I)$ is ``too smooth" given the characteristics of the
red spectrum.
{\co Note that $I = 100$ is close to $I_o$, {\it i.e.} the value that}
maximises the Kolmogorov--Smirnov probability in the simulations of sect.~3 
for red noise spectrum B.

It is clear from the discussion above that except for the first and last 
5--6 Fourier frequencies of each spectrum, the approximate $pdf$ derived above 
for $R_j(I_o)$ provides accurate results, as long as $I_o> 30-40$. 

\section{PERIODIC SIGNAL DETECTION}

The technique described above was developed in order
to approximate through a suitable smoothing 
the coloured noise components from the source variability and recover a white
noise sample spectrum in which the search for narrow peaks can be carried out
by applying the $pdf$ of the divided spectrum (sect.~4). In this section we
outline the procedure for the detection of periodic signals and 
the derivation of upper limits in case no a periodic signal is found.

Given a sample spectrum $P_j$, the divided spectrum $R_j(I) =2 P_j/S_j(I)$ is
calculated for a given smoothing width and its distribution compared to a  
$\chi^2_2$ $pdf$ by using a KS test. This is repeated for smoothing
widths ranging from a maximum of 
{\co twice} the number of Fourier frequencies in the  sample spectrum
to a minimum of $30-40$, with a spacing
of half an octave. The smoothing width $I_o$ that is found to produce the
highest KS probability is then adopted for the rest of the analysis. The
divided spectrum $R_j(I_o)$ is then searched for significant peaks testyfing
to the presence of a periodic modulation. The detection threshold is
determined by the expected distribution of the divided sample spectrum
$R_j(I_o)$, which is worked out based on the approximate $pdf$ of eq.~10. Note
that, unlike the case of a standard search in the presence of 
simple white noise,
the detection threshold depends on the trial frequency (cf. eq.~\ref{eq:pdfeq}).
By definition the threshold is given by the set of values $D_j(I_o)$ that
will not be exceeded by chance at any of the $J_{trial}$ frequencies examined,
with a  confidence level $C$. The (small) probability $1-C$ that
at least one of the $J_{trial}$ values of $R_j(I_o)$ exceeds the detection
threshold is given by 
\begin{eqnarray}
 1-C &=& 1 - [1 - Prob_{single}(R_j(I_o)>D_j(I_o))]^{J_{trial}}\\ \nonumber
&\simeq& J_{trial} Prob_{single}(R_j(I_o)>D_j(I_o)) ~,
\end{eqnarray}
where
\be
Prob_{single}(R_j(I_o)>D_j(I_o))
   = \int_{R_j(I_o)}^{+\infty} dr  \int_{-\infty}^{+\infty} |s| f_{P_j,S_j(I_o)}
   (rs,s) ds 
\ee
is the chance  probability of $R_j(I_o)$ exceeding the detection threshold
$D_j(I_o)$. By solving the equation
$Prob_{single}(R_j(I_o)>D_j(I_o)) \simeq (1-C)/J_{trial}$ 
for $D_j(I_o)$ , the detection threshold is obtained for each $j$.

The reliability of the whole procedure was tested by carrying out 1000
simulations in which 1024 frequency sample spectra with selected noise
components were searched for significant peaks with a confidence level of
$C=99\%$ in each sample spectrum (the search excluded
the first and last five Fourier frequencies).
This was repeated for four different types of sample spectra obtained from  
autoregressive processes: a white noise, a red noise, 
a white noise plus a broad peak and a red noise plus a broad peak. 
Fig.~8 gives an example of each of these spectra, together with the 
frequencies and chance probabilities (for $J_{trial}$=1014)of the peaks 
exceeding the detection threshold. Respectively $6$,
$11$, $15$ and $13$ peaks above the threshold were found, consistent with the 
expected value of $10$.

Once a peak $R_{j'}(I_o)$ above the detection threshold is found, the
corresponding signal power $P_{j',sig}$ is obtained {\co from the prescription 
of Groth (1975) and Vaughan et al. (1994) (see below). }
The sinusoidal amplitude, $A$, is
defined by assuming that the signal is given by $C_o~[1+A~sin(2\pi$
$\nu_{j',sig}t+\phi)]$, with $C_o$ the average count rate. $A$
is then derived by using standard methods; for binned data we have (see e.g. 
Leahy et al. 1983).
\be
  A = \left\{ \left[ \frac{P_{j',sig}}{2} \right] \frac{4}{0.773 N_{\gamma}}
  \frac{(\pi j/N)^2}{sin^2(\pi j/N)} \right\} ^\frac{1}{2} ~,
\ee
where $N$ is the number of bins in the light curve.

In the absence of a positive detection, the threshold $D_{j}$ must be converted
to an upper limit on the (sinusoidal) signal amplitude. To this aim it is
first necessary to calculate the sensitivity of the search, i.e. the
weakest signal $R_{j,sens}(I_o)$
that will produce, with confidence $C$, a
value of $R_j(I_o)$ exceeding the detection threshold $D_j(I_o)$.
This is done by using the prescription of Groth (1975) and Vaughan et al.
(1994) to calculate the  probability distribution of the total power resulting
from the signal and the (white) noise. Their procedure is adapted
to our case by using  $D_j(I_o)$ in place of their detection threshold for the 
power.
Strictly speaking the $pdf$ of the noise in the divided spectrum $R_j(I_o)$
substantially exceeds the $pdf$ of a simple $\chi^2_2$  distribution for high
values (see sect.~4 and fig. 4). However, the difference is only minor for
values of a few (2--3) standard deviations above the  average. The $pdf$
derived by Groth (1975) is therefore expected to apply to the noise in the
divided spectrum, for any reasonable value\footnote{
This is unlike the derivation of the detection threshold  
$D_j(I_o)$ which requires a detailed knowledge of the $pdf$ for
higher values, given the usually high
number of $J_{trial}$.} of the confidence level $C$. 
$R_{j,sens}(I_o)$ needs to be worked  out for each frequency $j$, because of
the $j$--dependence of $D_j(I_o)$. For the same reason it is not possible to
derive an upper limit which applies to all $j$'s, based on the highest
observed value of $R_j(I_o)$ as described by  van der Klis ({\co 1988}) and 
Vaughan et al. (1994).
The set of $R_{j,sens}(I_o)$ obtained in this way are then converted to
the corresponding powers through $P_{j,sens}(I_o)=R_{j,sens}(I_o)(S_j(I_o)/2)$ 
and then to the sinusoidal signal amplitude $A_{j,sens}$ by using eq.~13.
A similar procedure is used to derive confidence intervals
on the power $P_{j',sig}$, and therefore the amplitude $A$,
of a detected signal ($P_{j',sig}$ is used
in place of $D_j$ in this case).

\section{APPLICATION AND RESULTS}

In this section we apply the technique described above to
the light curves of two highly variable X--ray  binaries containing an
accreting neutron star. The data were obtained with  the Medium Energy
(ME) proportional counter array  on board the EXOSAT satellite (Peacock \& 
White 1988).  

V0332+53 is a 4.4~s X--ray pulsar accreting from a Be star in a transient 
X--ray binary. Its light curves are characterised by a pronounced red--noise 
type variability. 
The 4.4~s pulsations have a small amplitude ($A$$\simeq$8\%; Stella et al. 
1985 and references therein).   
We re--analysed a 4 hr long 1--9 keV light curve obtained on 1984 January 24 
with the EXOSAT ME, during which the source was 
relatively faint (average rate of $\sim$36 counts~s$^{-1}$). The light curve was
binned to a time resolution of 938~ms and a single sample spectrum calculated
over a 16384 point interval (see curve $a$ of fig.~9). 
A smoothing width of $I_o$=1803 Fourier frequencies
was obtained through the KS test described in sect.~3. 
In order to save on computer time, a preliminary search for narrows peaks is
carried out by using the rescaled $\chi^2_2$--like $pdf$ of $P_j$ in eq.~8, 
and modifying accordingly the procedure described in sect.~5   
(this is equivalent to ignoring the statistical uncertainties in the estimate 
of the smoothed spectrum $S_j(I_o)$). It is apparent from fig.~4 that a 
$\chi^2_2$ $pdf$ is systematically lower than the $pdf$ of $R_j(I_o)$  
$pdf$ in eq.~8 for high values. Therefore any divided spectrum peak
exceeding the detection threshold given by $D_j(I_o)$ will necessarily
correspond to a $P_j$ which exceeds the preliminary detection threshold
obtained from  eq.~8. A narrow peak at a frequency of  0.229 Hz  
(period of 4.376~s) was detected above the preliminary detection threshold
for a 3$\sigma$ confidence level (see curve $b$ in fig.~9).   
It is apparent from fig.~9 that the peak lies upon an extended  red--noise
component with an approximate power law slope of $-1$.  
The probability of chance occurrence of the corresponding peak 
in $R_j(I_o)$ was then worked out on the  basis of the $pdf$ in eq.~10
and the prescription in sect.~5. 
This gave a  probability of chance occurrence of 8.6$\times$10$^{-5}$ for 
8182 trial frequencies. Therefore the technique that we have developed 
would have led to the discovery of the $\sim 4.4$~s pulsation period 
based only on the spectrum of fig.~9. 
The pulsation amplitude was determined to be $7$$\pm$$1$ ($1$$\sigma$ error
bars) consistent with the values quoted in the literature. 

GX$13$$+$$1$ is a relatively poorly studied bright galactic bulge X--ray
source, which displays only aperiodic variability up to frequencies of 
{\co several tens of Hz (Garcia et al. 1988; Hasinger \& van der Klis 1989)}. 
The source is likely a binary system containing an old neutron star accreting
from a low mass companion. 
We have reanalysed the 1985 EXOSAT ME observations of GX13+1, searching for 
a periodic modulation of the X--ray flux in the frequency domain 
shortwards of $10^{-2}$~Hz where the red--noise variability is clearly 
dominant. 
For the $\sim$~6 hr observation of 1985 April 1 and the $\sim$~7 hr 
observation of 1985 May 2, light curves with a resolution 
of 2.5~s and 1~s, respectively, were accumulated in different energy
bands (1--4, 4--9 and 1--9~keV). The power spectra were calculated over the 
longest possible interval
for each observation in order to extend the search to  the lowest sampled
frequencies.  The self--consistent determination of the smoothing width
provided values of $I_o$ between $\sim$$67$ and $\sim$$1422$ Fourier
frequencies for different energy ranges. 
No significant peaks above the $95\%$ confidence detection threshold
were found in any of the power spectra. The corresponding upper limits to the
amplitude of a sinusoidal modulation were worked out for each observation 
in the three different energy intervals. Two examples are given by  curves $c$
in fig.~10. Values for selected trial periods are listed in Table 1; they 
range  from $\sim$$1\%$ to $30\%$ for periods between $200$ and 
$10000$~s.
The behaviour of these upper limits clearly reflects the 
shape of the red--noise component of the sample spectra, such that  
source variability, rather than counting statistics noise, 
limits the sensitivity of the period searches. 

\section{CONCLUSION}

The power spectrum analysis technique that we developed for the detection
of periodic signals in the presence of ``coloured" noise components 
presents the following main advantages: 
$(i)$ it does not require any reduction of the Fourier frequency resolution;
$(ii)$ it can be used also in the presence of the relatively steep 
red noise components (power law slopes as low as --2) which are commonly
found in nature; 
$(iii)$ it takes into account the statistical uncertainties in the 
estimator of the continuum power spectrum components. 
Extensive numerical simulations were carried out in order to test the
reliability of the technique and define the range of applicability of the 
adopted approximations. We found that very good results are obtained if the
first and the last 5--6 Fourier frequencies of the sample spectrum are 
excluded from the analysis and the width of the smoothing is larger
than 30--40 Fourier frequencies. 

Though based on relatively simple statistics, the numerical evaluation of the 
peak detection threshold must be carried out separately for each 
Fourier frequency: this can be quite CPU--intensive. A way around this
limitation consists in carrying out a much faster
search for potentially significant peaks by using a preliminary and simpler 
detection threshold.
The significance of candidate peaks is then reassessed on the basis of
a complete application of the technique described in this paper. 
Computer programs based the technique described in this paper will be made
available to the community through the timing analysis package {\it Xronos}
(Stella \& Angelini 1992a,b).

\acknowledgments

The authors acknowledge useful discussions with L. Angelini and A. Parmar. 
{\co M. van der Klis provided useful comments on an earlier version of the 
paper.} 
This work was partially supported through ASI grants.


\appendix
\section{GENERALIZATION TO $M$ ($>$1) SPECTRA}

This appendix describes the more general case in which the sample spectrum is 
calculated by summing $M$ ($>$1) individual sample spectra each obtained from a 
separate light curve interval. 

The normalisation of the power estimates in eq.~4, is converted to   
\be
P_{j}=\sum_{i=1}^{M} P_{j,i} = \sum_{i=1}^{M} \frac{2}{N_{\gamma,i}}
|a_{j,i}|^{2}~,~~~~~~~
\sum_{i=1}^{M} N_{\gamma,i}=N_{\gamma}~~~~~.  \nonumber
\ee
Being each of the 
$M$ variables $P_{j,i}$ distributed like a rescaled $\chi^2_2$, their sum 
$P_{j}$ will follow  
a $\chi^2$ distribution with 2$M$ $dof$, rescaled by the same factor.
The $pdf$ of $P_j$  (cf. eq.~8) is therefore given by 
\be
    f_{P_j}(p)=\displaystyle{\frac {1}{\Gamma(M)}
             \left( \frac{1}{2} \right)^{M} p^{M -1}
             (S_j(I_o))^{- M}
             e^{-p/2 S_j(I_o)}~(p)} ~~~~~.
\ee
The estimate of $|H(\nu_j)|^2$ is given by $S_j(I_o)/2M$ 
and consequently the divided spectrum is 
$R_j(I_o)$= 2$M$$P_j/S_j(I_o)$. The approximate 
$pdf$ of $R_j(I_o)$  given in eq.~10  converts to 
\begin{eqnarray}
   f_{R_j(I_o)}(r) &=& \frac{1}{\sigma \sqrt{2\pi}}
   ~\frac{2^{- M}}{\Gamma(M)}
   \cdot \int_{0}^{+\infty} s(rs)^{M-1}~
   (S_j(I_o))^{- M} \\ \nonumber
   ~& &~~exp \left[- \frac{(s-\mu)^{2}}{2\sigma^{2}} -
   \frac{rs}{2 S_j(I_o)} \right] ds ~~~~~. 
\end{eqnarray}
This approximate $pdf$ remains fairly accurate for $M I_o\geq 40$.
Finally the sinusoidal signal amplitude (eq.~13) is given by   
\be
  A = \left\{ \left[ \frac{P_{j',sig}}{2M} \right] \frac{4}{0.773 N_{\gamma}}
    \frac{(\pi j/N)^2}{sin^2(\pi j/N)} \right\} ^\frac{1}{2} ~~.
\ee
In order to 
test the reliability of the technique, we also carried out 
extensive numerical simulations for a variety of cases in which $M>1$. 

\newpage
\begin{table}
\begin{center}
\begin{tabular}{lcccccc}
\multicolumn{7}{c}{TABLE ~1}\\
\multicolumn{7}{c}{GX$13$$+$$1$ } \\
\tableline \tableline 
Period& \multicolumn{6}{c}{95\% confidence upper limits on A}\\
$s$&\multicolumn{3}{c}{1985 April}& \multicolumn{3}{c}{1985 May} \\
\cline{2-7}
 &1--9&1--4&4--9&1--9&1--4&4--9\\
\tableline
10240& 6.3\% & 11.1\% & 3.7\% & 20.1\% & 29.6\% & 12.4\% \\
6830 & 5.6   &  9.1   & 6.2   & 14.5   & 21.1   &  9.2  \\
5120 & 6.2   &  8.6   & 5.9   & 10.0   & 14.7   &  6.3  \\
4096 & 4.7   &  6.5   & 6.0   &  8.3   & 12.0   &  5.4  \\
2048 & 2.4   &  3.3   & 3.0   &  4.7   &  6.9   &  2.8  \\
1024 & 1.4   &  2.3   & 1.8   &  2.5   &  3.7   &  1.7  \\
256  & 0.8   &  1.3   & 0.9   &  1.1   &  1.4   &  1.0  \\
128  & 0.7   &  0.9   & 0.8   &  0.7   &  0.9   &  0.8  \\
\tableline
\end{tabular}
\end{center}
\end{table}

\newpage

\section*{ Figure captions}
 
{\bf Figure~1:} (upper panel) Average sample spectrum from a 1-20~keV EXOSAT 
observation of the black hole candidate X-ray binary Cyg~X-1. (lower panels)
Distribution of the normalised spectral estimates for selected Fourier
frequencies ($j= 6, 10, 20, 40, 60$, see arrows in fig. 1a). The solid lines
represent a $\chi^2_2$-distribution. 

\vspace{4mm}

{\bf Figure~2:} Comparison between the standard rectangular smoothing technique
(dashed lines) and the logarithmic interval smoothing technique (solid lines)
for the {\co three} ``coloured" spectral shapes discussed in the text. Lines
and points represent the average from 1000 simulations.
The smoothing width is $I=30$.
The difference between the two smoothing techniques at a single Fourier
frequency is sketched in the {\co central} panel (the numbers indicate
$I_{left}$ and $I_{right}$, i.e. the number of Fourier frequencies
used at the left and the right of the nominal frequency $\nu_j$).

\vspace{4mm}

{\bf Figure~3:} Kolmogorov--Smirnov probability as a function of the smoothing
width $I$ for the comparison between a $\chi^2_2$--distribution and
$R_j(I)$ in {\co three} different cases: a white noise spectrum (W)
and the ``coloured" spectra of fig.~2 (A{\co , B and C}).

\vspace{4mm}

{\bf Figure~4:} Distribution of $R_j(I)$ from $2\times10^4$ simulations of a
5000 frequency white noise spectrum, for selected values of the smoothing width
$I$. Note that only the power estimates between j=6 and j=4995 were considered.
The lines give the expected $pdf$, calculated as described
in the text. For comparison a $chi^2_2$ $pdf$ is also shown at the bottom of 
the figure.

\vspace{4mm}

{\bf Figure~5:} Distribution of $R_j(I)$ from $10^6$ simulations of a
1000 frequency white noise spectrum, for $I_o$=100 and selected Fourier
frequencies close to the low-frequency end ($j$=5,6,7,10). 
The lines give the expected $pdf$, calculated as described
in the text.

\vspace{4mm}

{\bf Figure~6:} As fig.~5 but for the last frequencies of spectrum 
close to the Nyquist frequency ($j$=995,994,993,990).

\vspace{4mm}

{\bf Figure~7:} Distribution of $R_j(I)$ from $2\times10^4$ simulations of a
5000 frequency red noise spectrum, for selected smoothing widths
$I$. Only the power estimates
between j=6 and j=4995 were considered.

\vspace{4mm}
{\bf Figure~8:} Left panels: examples of the sample spectra obtained from
autoregressive processes and searched for coherent pulsations (10$^3$  
simulations). From top to bottom the power spectra comprise: 
a red noise, a white noise, a white noise plus a broad peak 
centered at 0.15 Hz and a red noise plus a broad peak centered at 0.25 Hz 
Right panels: probability of the peaks exceeding the 99\% confidence 
preliminary detection threshold (see sect. 6). 
The probability is calculated from the $pdf$ of $R_j(I_o)$ in eq.~10. 
From top to bottom only 
6 11, 15 and 13 peaks, respectively, were found 
to exceed to 99\% confidence threshold (shown by the horizontal line).
This is consistent with the expected value of 10. Note that in the presence 
of red noise most of the peaks detected with the preliminary threshold 
turn out not be be significant. 

\vspace{4mm}

{\bf Figure~9:} Power spectrum of the 1--9~keV EXOSAT ME light curve of
X--ray transient source V$0332$$+$$53$ obtained during the 1984 January 24
observation (a), together with the preliminary threshold for the detection
of sinusoidal periodicities at 3$\sigma$ confidence level (b). The first and
last five frequencies of the power spectrum are not included in the search
for periodicities.

\vspace{4mm}

{\bf Figure~10:}  Power spectra of the 1--9~keV EXOSAT ME light curves of GX$13$
$+$$1$ for the 1985 April 1(upper panel) and May 2(lower panel) observations 
(a), together with the preliminary threshold for the detection of sinusoidal 
periodicities at the 95\% confidence level (b). Curve (c) shows the 
corresponding upper limits on the pulsation amplitude $A$.

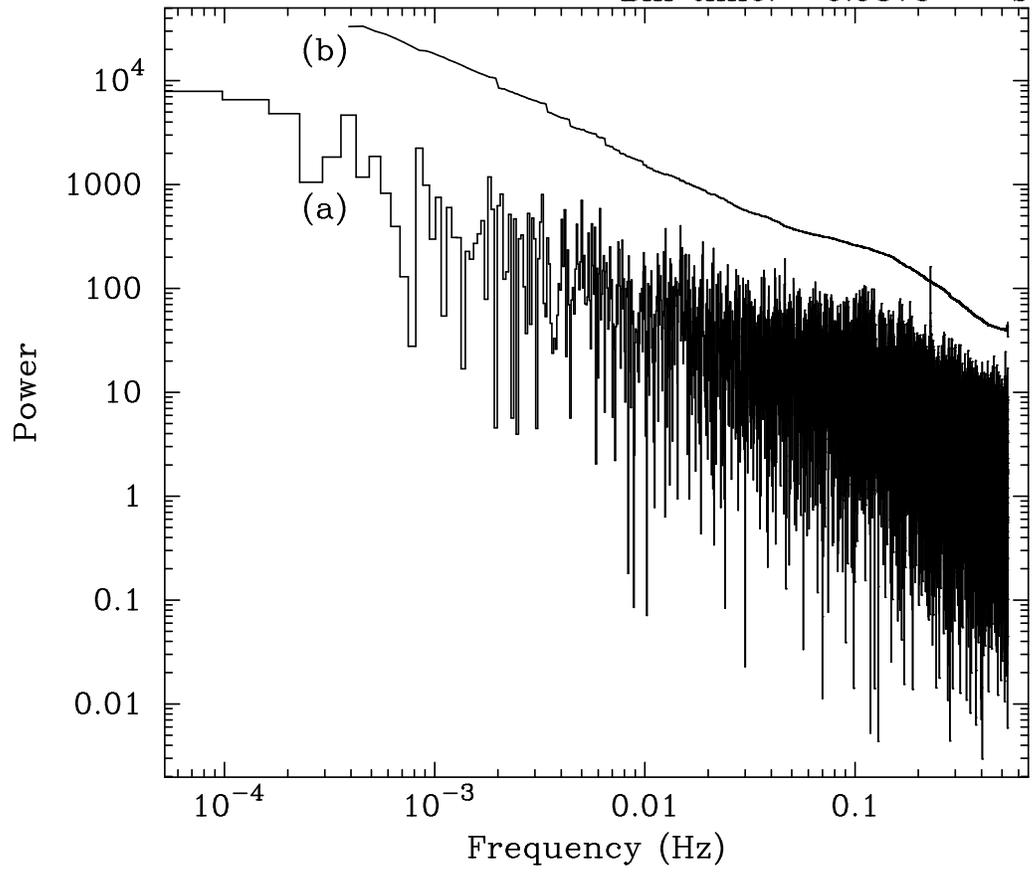
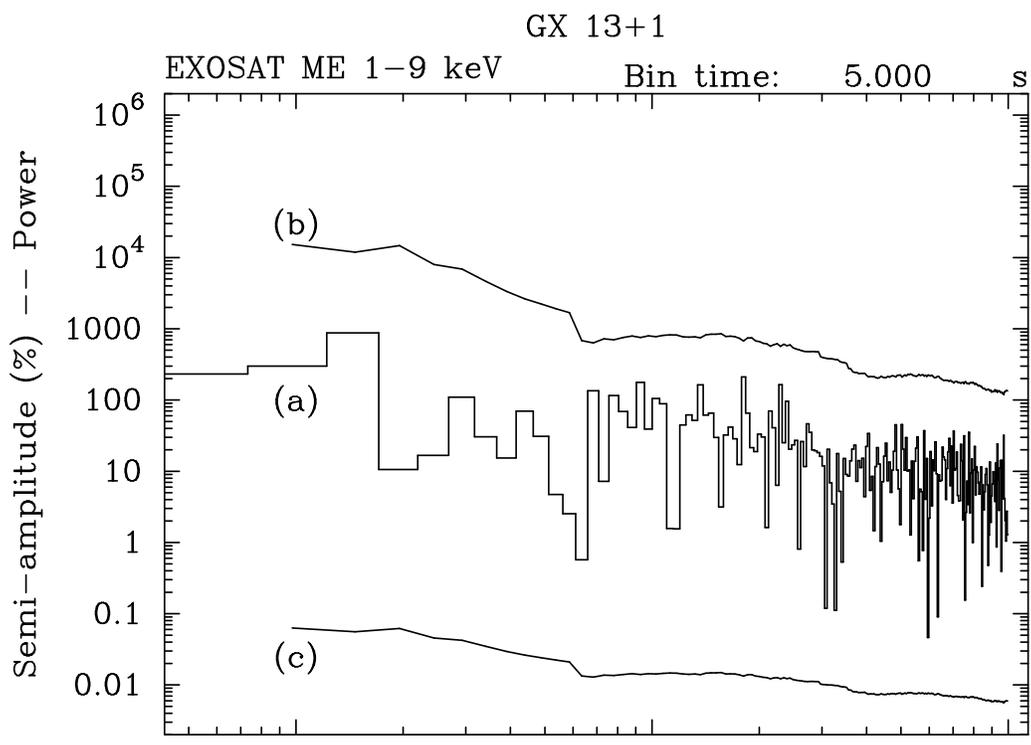
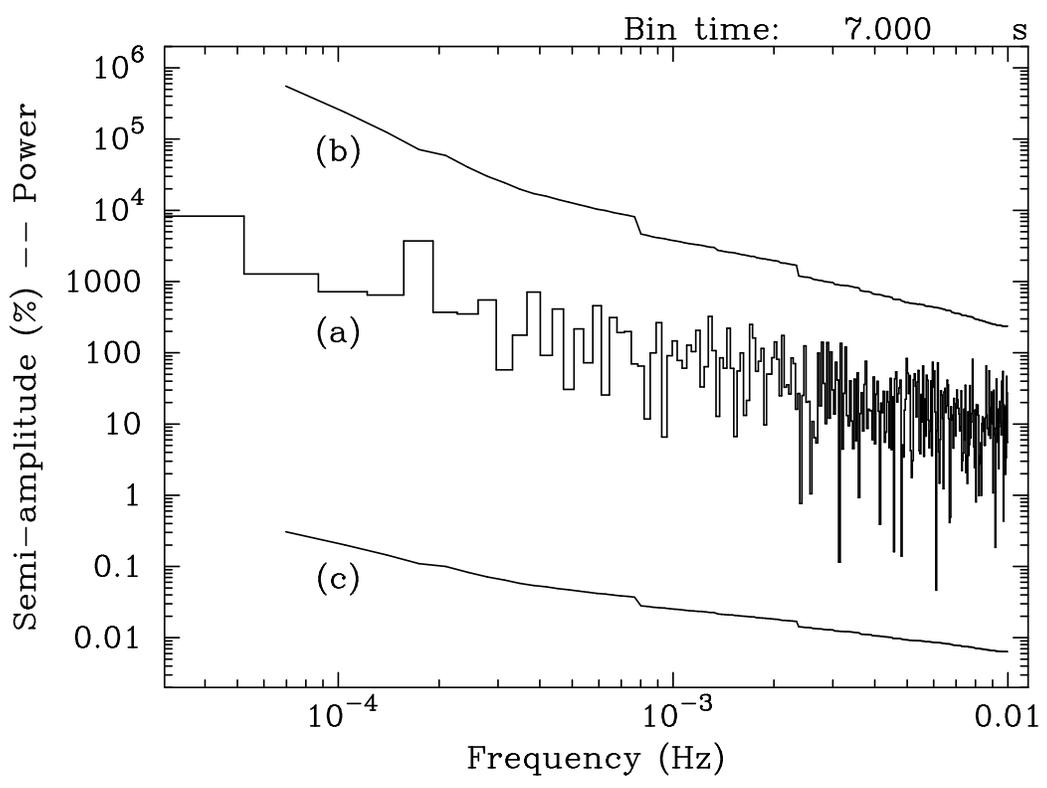
\end{document}